\documentclass[preprint,12pt]{aastex}
\usepackage{times}
\usepackage{epsfig}
\usepackage{natbib}

\begin{document}

\title{A NOTE ON THE TORSIONAL OSCILLATION AT SOLAR MINIMUM}

\author{
R.~Howe,\altaffilmark{1}\email{rhowe@noao.edu} 
J.~Christensen-Dalsgaard,\altaffilmark{2} 
F.~Hill,\altaffilmark{1} R.~Komm\altaffilmark{1}, 
J.~Schou,\altaffilmark{3} 
and M.~J.~Thompson\altaffilmark{4}
}

\altaffiltext{1} {National Solar Observatory, P.O. Box 26732, Tucson AZ 85726-6732, USA}
\altaffiltext{2} {Department of Physics and Astronomy,
Aarhus University, DK-8000 Aarhus C, Denmark}
\altaffiltext{3}{
HEPL Solar Physics,
452 Lomita Mall,
Stanford University,
Stanford, CA 94305-4085, USA}
\altaffiltext{4}{
School of Mathematics and Statistics,
University of Sheffield,
Hounsfield Road,
Sheffield S3 7RH,
UK}

\begin{abstract}
We examine the evolution of the zonal flow pattern in the upper solar
convection zone during the current extended solar minimum, and compare it 
with that during the previous minimum. The results suggest that 
a configuration matching that at the previous minimum was reached during 2008, but that the flow band corresponding to the new
cycle has been moving more slowly towards the equator than was observed
in the previous cycle, resulting in a gradual increase in the apparent
length of the cycle during the 2007\,--\,2008 period. 
The current position of the lower-latitude fast-rotating belt 
corresponds to that seen around the onset of activity
in the previous cycle.
\end{abstract}


\section{INTRODUCTION}
The onset of Solar Cycle 24 appears to be later than expected; nearly thirteen years after the Cycle 23 minimum in mid-1996 there is still very little surface magnetic activity.
In this Letter we examine the large-scale zonal flow pattern 
seen in the upper convection zone in recent observations and compare the results with those for the previous solar minimum.

The migrating zonal flow pattern known as the
torsional oscillation was first detected by \citet{1980ApJ...239L..33H} in
Doppler measurements at the solar surface carried out at the Mount Wilson Observatory; it consists of belts
of slightly faster than average rotation that migrate from mid-latitudes
to the equator and poles. The migration of the zonal flow bands during the solar cycle is closely connected to the migration of the magnetic activity belt. 
The 
flow pattern was first detected helioseismically by \citet{1997ApJ...482L.207K} in early $f$-mode data from the Michelson Doppler Imager (MDI) aboard the Solar and Heliospheric Observatory spacecraft, 
and was seen to 
migrate by \citet{1999ApJ...523L.181S}. Using $p$-mode data it is possible to resolve the depth 
penetration of the flows; however, only the North\,--\,South symmetric
part of the pattern can be detected using global-mode helioseismology.
The first four/five years of zonal-flow results from Global Oscillation Network Group (GONG) and MDI $p$-mode and $f$-mode data were reported by \citet{2000ApJ...533L.163H}, while \citet{2001ApJ...559L..67A} and \citet{2003safd.book..247S} drew attention to the 
high-latitude part of the pattern. The further evolution of the pattern in 
Cycle 23 was described by \citet{2002Sci...296..101V,2003ApJ...585..553B,2005ApJ...634.1405H}, and by \citet{2006SoPh..235....1H}, who also considered the 
updated Doppler observations from Mount Wilson. The analysis of 
\citet{2008ApJ...681..680A} extends these investigations into the current solar minimum.
It became evident that the 
torsional oscillation pattern penetrates through a significant 
fraction of the convection zone; 
\citet{2005ApJ...634.1405H} also found 
some phase variation with depth in the lower-latitude, equatorward-moving part 
of the pattern, in the sense that the increase in rotation rate 
happens at earlier times at deeper depths. The lower-latitude branch of the
pattern migrates equatorward with the activity belts, starting at about $45\deg$ latitude a few years before the previous solar minimum and reaching the equator a couple of years after maximum; the broader, stronger belt poleward of $45\deg$ propagates poleward on a rather shorter timescale.  The whole phenomenon 
is usually believed to be a side-effect of the dynamo that drives the solar cycle, either directly through the Lorentz effect \citep{1981A&A....94L..17S} or 
indirectly through geostrophic effects in the activity belt \citep{2003SoPh..213....1S}. Nevertheless, the bands are clearly detectable even when there are 
few or no organized active regions, as was seen in the early years of Cycle 23 and again at the dawn of Cycle 24. The magnetic or sunspot butterfly diagram
does not exhibit purely periodic behavior; similarly, there is no reason to expect that the zonal flow pattern should repeat precisely from cycle to cycle, but it is interesting to compare the flows seen at different minima and the 
way they evolve.

In this Letter, 
we consider the behavior of the flows around the current and 
previous solar minima. 
We concentrate on the equatorward-moving part, basing our comparisons 
on latitudes equatorwards of $45\deg$, and on the
upper part of the convection zone where the signal is clearest.

\section{DATA}

We now have inferences of the solar interior rotation rate,
$\Omega(r,\theta$) 
where $r$ is radius and $\theta$ is latitude, covering the period 
from mid-1995 (GONG) and mid-1996 (MDI) to early 2009. 
These inferences were obtained 
using 2-dimensional regularized least-squares inversions on sets of medium-degree $p$-mode 
data from a total of 
137 GONG time series (108 days, with start times at 36-day intervals)
and 64 non-overlapping MDI time series of 72 days each, contiguous 
except for the interruptions due to SOHO problems in 1998 and 1999. 
Details of the
data analysis and inversion techniques are given in \citet{2005ApJ...634.1405H} and references therein. 
The inversions are carried out on a mesh with 48 equal intervals in latitude and 50 non-uniform intervals in radius; the radial mesh was chosen to be 
approximately uniform in acoustical radius. 
To reveal the torsional oscillation signal, at each location a
temporal mean over all epochs was subtracted from the rotation-rate inference.
The mean was calculated separately for MDI and GONG inversions, before the residuals were combined into one time series.

The synoptic magnetic index used here has been derived from a combination of KPVT\footnote{Available from {\tt http://nsokp.nso.edu/dataarch.html}} and 
SOLIS VSM \footnote{Available from {\tt http://solis.nso.edu}} magnetograms.

\section{RESULTS AND ANALYSIS}
\begin{figure}
\plotone{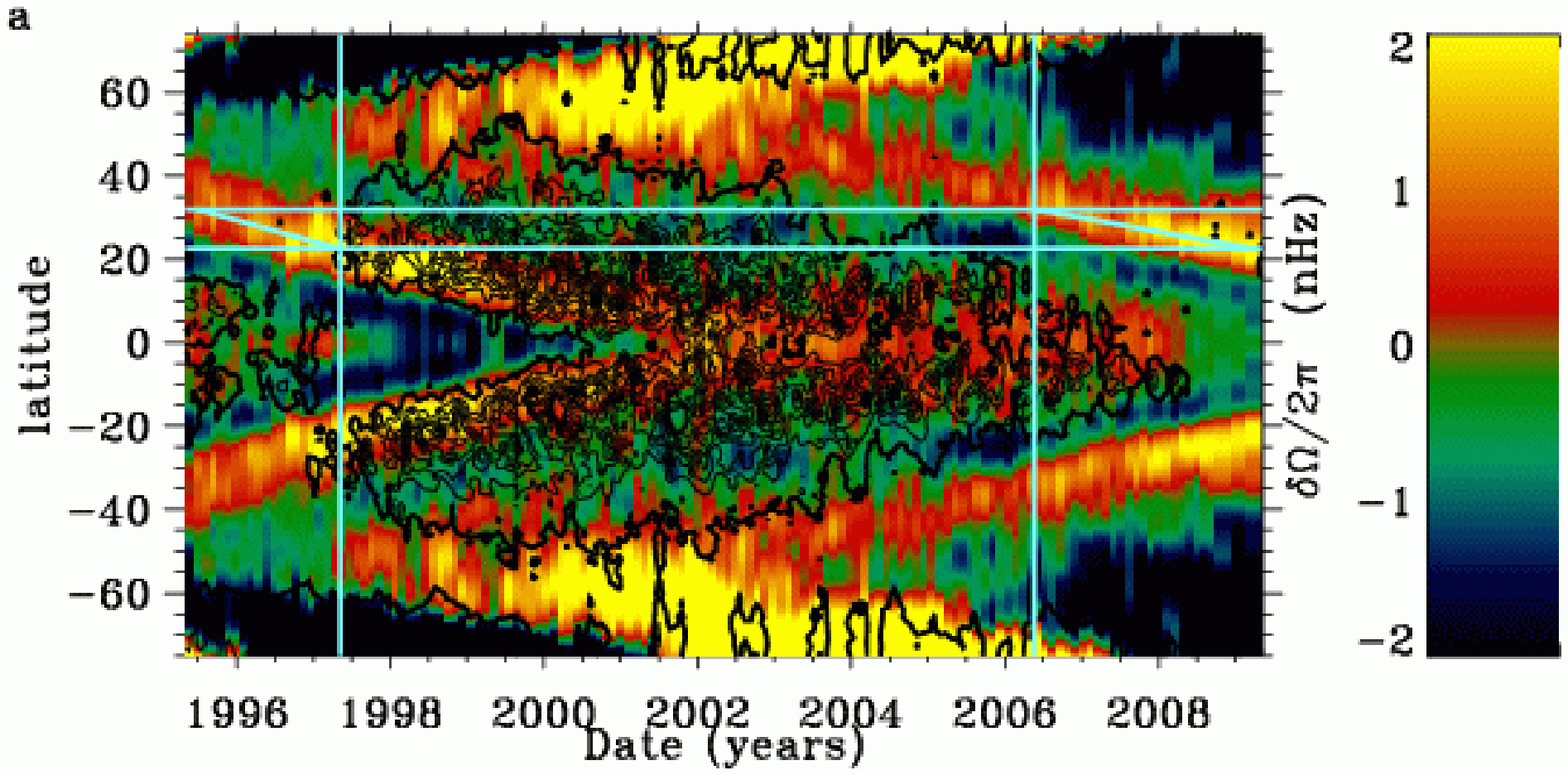}
\plotone{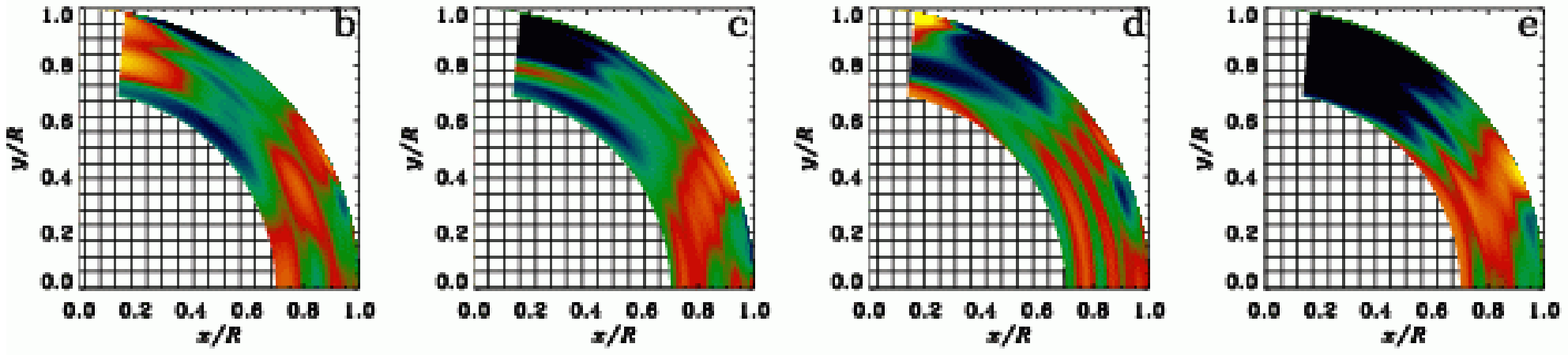}
\caption{\label{fig:fig1}({\it a}) Rotation-rate residuals at $0.99 R_\odot$ from 
MDI and GONG. Overlaid contours show the gross longitudinal magnetic
field strength from KPVT/SOLIS, at 5G intervals. The leftmost solid vertical light-blue
line shows the date, 1997.3, at which the low-latitude flow configuration
best matches that in the most recent (2009.2) data set, and rightmost vertical line the date, 2006.4,
where it best matches that in the earliest data set (1996.5), while the
horizontal lines show the respective location of the flow bands and the 
slanted lines schematically indicate the migration of the equatorward branch.
The lower panels show 12-month averages of the rotation-rate residuals in the $r,\theta$ plane for epochs starting at ({\it b}) 1995.5, ({\it c}) 1996.3, ({\it d}) 2006.5, ({\it e}) 2008.2.}
\end{figure}

\begin{figure}
\plotone{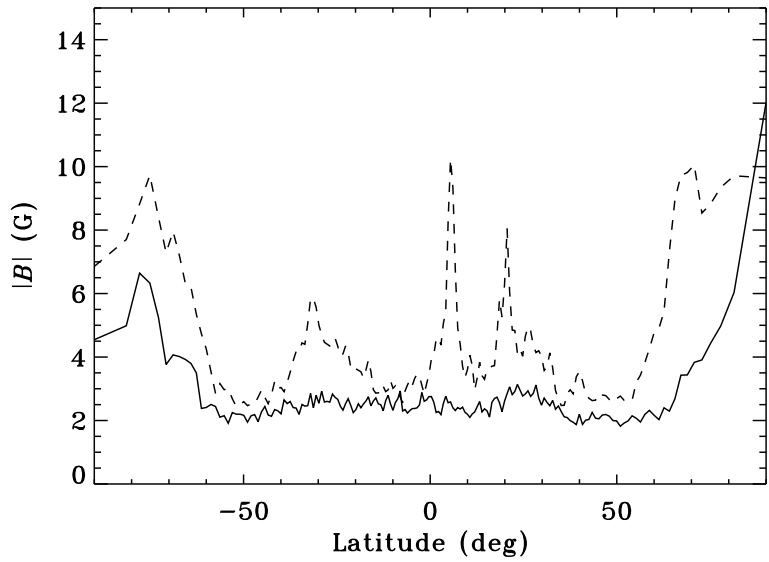}
\caption{\label{fig:fig2}KPVT/SOLIS gross longitudinal magnetic field strength as a function of latitude, 
for the 2009.2 epoch of the most recent zonal flow observations (solid line) 
and for the epoch, 1997.3, during the previous minimum where the flows most closely
match the most recent ones (dashed line).}
\end{figure}

Figure~\ref{fig:fig1} shows the zonal flow pattern
as a function of latitude and time at $0.99R_\odot$. 
The most recent MDI observations processed are from early 2009. Overlaid are contours
of the gross magnetic longitudinal magnetic field strength. The
branch of the torsional oscillation pattern corresponding to 
Cycle 24 is well established and clearly visible, but is moving more
slowly towards the equator than did the corresponding branch in Cycle 23.
The 
low-latitude near-surface flow profile in this data set is best matched by that at
date 1997.3, as shown by the left-hand vertical line in the figure. At this point in 
cycle 23, substantial magnetic activity was just starting, as
illustrated by the latitudinal magnetic profiles in Figure~\ref{fig:fig2}.
The level of activity seen for the current cycle is much lower, but what 
hints of new-cycle activity there are seem to be at the correct latitude. Note also that
there seem to have still been a few old-cycle active regions around at this point in early 1997; at present only a few very weak old-cycle regions are seen, for example, NOAA 11016 on 2009 April 29 --- May 1.
To give an idea of the depth dependence of the flows, Figure~\ref{fig:fig1} also shows twelve-month averages of the residuals in the $r,\theta$ plane for 
four selected low-activity epochs. The radial detail of these profiles should
not be over-interpreted, due to the poorer signal-to-noise ratio of the
inversions in the interior and the varying resolution of different inversions, but the penetration of the flow belt into the bulk of the convection zone 
is clear.

\begin{figure}
\plotone{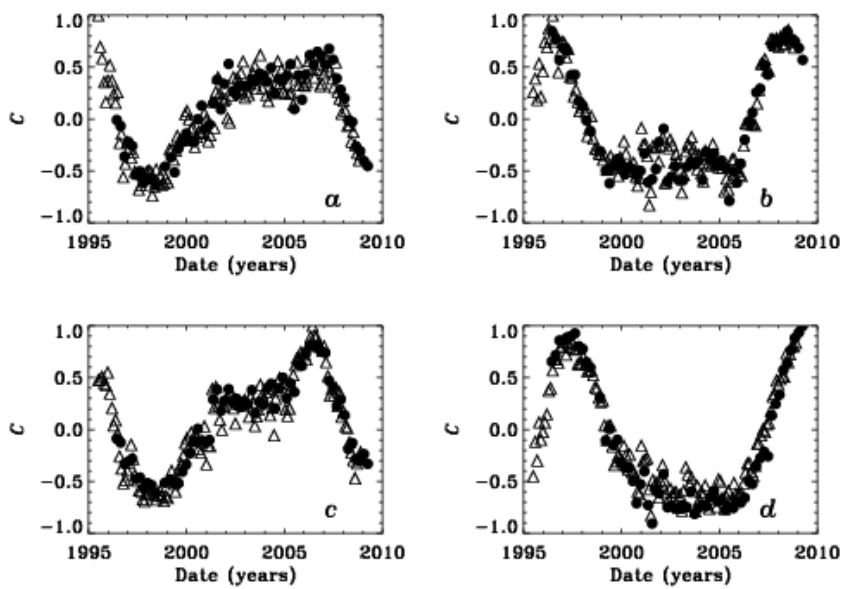}
\caption{\label{fig:fig4}Correlation between near-surface rotation
residuals at each time sample and those at selected reference epochs. Triangles represent GONG data and circles MDI.
The reference epochs are: ({\it a}) one of the earliest sets of GONG observations at
1995.5; 
({\it b}) the solar minimum at 1996.5; 
({\it c}) the epoch 2006.5, 
chosen to give the best match to the earliest observations; 
and ({\it d}) the last set of observations we have at 2009.2. 
}
\end{figure}

How does the evolution of the flows during the current minimum correspond to
what was seen in the previous one?
To quantify this, we consider  the linear correlation between the near-surface residuals $\delta\Omega(\theta,r,t_1)$ at each time step $t_1$ and those at a reference epoch $t_0$, $\delta\Omega(\theta,r,t_0)$. The correlation coefficient between two variables $x(t_0)$ and $x(t_1)$ is defined \citep{Bevington} as 
\begin{equation}
{C(t_0,t_1)}={{{\sum_i{[x_i(t_0)-{\bar{x}(t_1)}][x_i(t_1)-{\bar{x}(t_1)}]}}}\over
{\{\sum_i[x_i(t_0)-{\bar{x}(t_0)]^2\}^{1/2}}}\{\sum_i[x_i(t_1)-{\bar{x}(t_1)]^2\}^{1/2}}}\label{eq:eq1}\;,
\end{equation}
where in this case $x_i(t)\equiv\delta\Omega(\mathbf{r}_i,t)$ and 
the index $i$ enumerates inversion mesh points~{$\mathbf{r}_i$} with 
radius $0.97 \leq r/R_\odot \leq 0.995$ and 
$0\leq \theta \leq 45\deg$.
In Figure~\ref{fig:fig4} we show the results of this analysis for 
four selected reference epochs.
For a reference epoch of $1995.5$ (Figure~\ref{fig:fig4}{\it a}), about a year before the last solar minimum, the correlation reaches a maximum value around the second half of 2006 to early 2007 and then declines sharply. 
When the reference epoch is at 1996.5, corresponding to the previous solar minimum, (Figure~\ref{fig:fig4}{\it b}) 
the correlation coefficient reaches a maximum around the middle of  
2008, and then starts to decline.
A reference epoch of 2006.5 (Figure~\ref{fig:fig4}{\it c}) gives a maximum correlation at the beginning of the observations, 1995.5, while a reference epoch of the most recent set of observations, 2009.2, gives a correlation peaking at 1997.3, about eleven months after solar minimum (Figure~\ref{fig:fig4}{\it d}). 

\begin{figure}
\plotone{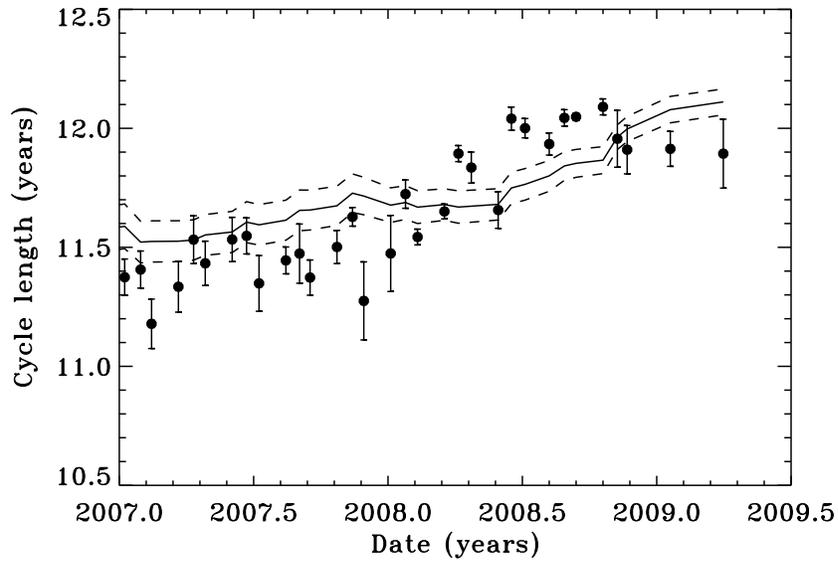}
\caption{\label{fig:fig3}
Estimated length of cycle as a function of last observing time,
averaged over $0.97 \leq r/R_\odot \leq 0.995$ and latitude less than $45\deg$.
Results are derived from time lag for best correlation (symbols, with 
the error bar representing the standard deviation of the mean), and length of cycle
estimated by sinusoid fit (solid line, with standard deviation of the mean shown 
by the dotted lines).}
\end{figure}

To study the temporal evolution of the apparent cycle length, 
for each data set from the beginning of 2007 onwards
we evaluate equation~(\ref{eq:eq1}) over latitudinal mesh points up to $45 \deg$, but separately
at each radial mesh point from $0.973$ to $0.995R_\odot$,
and find the date of the highest correlation value in the early part of the cycle. The difference between this date and the reference epoch gives an
estimate of cycle length in years;
we then take the mean and standard deviation of these lengths at each time step, thus obtaining an estimate of the effective cycle length and its uncertainty as a function of time.
The results, seen in Figure~\ref{fig:fig3},
seem to suggest that the effective length of the cycle measured in this manner has been slowly 
increasing through 2007\,--\,2008.

An alternative approach to describing the temporal variation of the flows, 
introduced by \citet{2002Sci...296..101V}, is to 
fit the variation at each location with a sinusoid function. 
\citet{2005ApJ...634.1405H} showed that, for the data through 2004, 
the variation was well described by fitting with an 11-year period sine wave and its second
harmonic, but did not find it practical to investigate other possible 
cycle lengths with only nine years of observations. With nearly fourteen years of data, we can now test the fit with other cycle lengths.
The time variation of the rotation at each location is modeled as 
\begin{equation}
\Omega(r,\theta,t)=\Omega_0(r,\theta)+A_1\sin(2\pi t/P)+
A_2\cos(2\pi t/P)+A_3\sin(4\pi t/P)+A_4\cos(4\pi t/P)\;,
\end{equation}
where $P$ is the period in years and time $t$ is also measured in years.
We then vary $P$ over the range from 9 to 14 years and find the value
which gives the lowest $\chi^2$ at each mesh point, before taking the mean
result
and its standard deviation over the range $0.97 \leq r/R_\odot \leq 0.995$.
Again, we see a tendency to gradually increasing perceived cycle 
length in these best-fit periods, as shown by the curve in Figure~\ref{fig:fig3}.  Taking these two analyses together, we estimate that the length of the
cycle in the region equatorwards of $45\deg$  
increased from about 11.2 to 12.2 years between early 2007 and early 2009.
We have carried out a similar sine-wave analysis at higher latitudes; poleward of $60\deg$ the cycle length remains almost constant at about 12 years over the same
period, while at intermediate latitudes the behavior is more complicated and
not clearly periodic.
In other words, the 
flow configuration, in agreement with other indicators, seems to indicate a prolonged minimum, with the flow belts (and, presumably, the almost-undetectable
magnetic activity belts) moving more slowly towards the equator than was
the case during the previous minimum.

\section{CONCLUSION}
The zonal flow pattern during the current solar minimum,
considered alongside that for the previous minimum and the 
magnetic butterfly diagram, suggests that the minimum was reached
during 2008, giving a length of approximately 12 years for Cycle 23. The flow band associated with the new cycle has been 
moving more slowly towards the equator than that observed during the
previous minimum. 

It would not be surprising to see a rapid increase in 
activity in the near future; it is tempting 
to suggest that if the current low levels persist for much longer we may indeed be looking at an unusually weak cycle.

In follow-up work, we will investigate the behavior of the flows in more detail, including the
poleward branch and the flow in the deeper layers.

\acknowledgments{This work utilizes data obtained by the Global Oscillation Network
Group (GONG) program, managed by the National Solar Observatory, which is
operated by AURA, Inc. under a cooperative agreement with the National
Science Foundation.
The data were acquired by instruments operated by the Big Bear Solar
Observatory, High Altitude Observatory, Learmonth Solar Observatory,
Udaipur Solar Observatory, Instituto de Astrof\'{\i}sica de Canarias, and
Cerro Tololo Interamerican Observatory.
The Solar Oscillations Investigation (SOI) involving 
MDI is supported by NASA grant NAG 5-13261
to Stanford University.  {\it SOHO} is a mission of international cooperation
between ESA and NASA.   
RK, and RH in part, were supported by NASA
contracts S-92698-F, NAG 5-11703, and NNG 05HL41I.
NSO/Kitt Peak data used here 
were produced cooperatively by NSF/NOAO, NASA/GSFC, and NOAA/SEL;
SOLIS data are produced cooperatively by NSF/NSO and NASA/LWS.
}

\newpage 
\epsscale{0.7}

\newpage

\newpage


\newpage

\end{document}